\documentclass[10pt]{article}
\usepackage{amsmath}
\usepackage{amssymb}
\usepackage{graphicx}
\usepackage{cite}
\usepackage{color}
\usepackage[labelfont=bf,labelsep=period,justification=raggedright]{caption}
\makeatletter
\renewcommand{\@biblabel}[1]{\quad#1.}
\makeatother

\date{}

\begin{document}

\begin{flushleft}
{\Large
\textbf{Synchronous bursts on scale-free neuronal networks with attractive and repulsive coupling}
}\sffamily
\\[3mm]
\textbf{Qingyun Wang,$^{1}$ Guanrong Chen,$^{2}$ Matja{\v z} Perc,$^{3,\ast}$}
\\[3mm]
{\bf 1} Department of Dynamics and Control, Beihang University, Beijing, China\\
{\bf 2} Department of Electronic Engineering, City University of Hong Kong, Hong Kong SAR, China\\
{\bf 3} Department of Physics, Faculty of Natural Sciences and Mathematics, University of Maribor, Slovenia\\
$\ast$matjaz.perc@uni-mb.si [www.matjazperc.com]

\end{flushleft}

\sffamily
\section*{Abstract}
This paper investigates the dependence of synchronization transitions of bursting oscillations on the information transmission delay over scale-free neuronal networks
with attractive and repulsive coupling. It is shown that for both types of coupling, the delay always plays a subtle role in either promoting or impairing synchronization. In particular, depending on the inherent oscillation period of individual neurons, regions of irregular and regular propagating excitatory fronts appear intermittently as the delay increases. These delay-induced synchronization transitions are manifested as well-expressed minima in the measure for spatiotemporal synchrony. For attractive coupling, the minima appear at every integer multiple of the average oscillation period, while for the repulsive coupling, they appear at every odd multiple of the half of the average oscillation period. The obtained results are robust to the variations of the dynamics of individual neurons, the system size, and the neuronal firing type. Hence, they can be used to characterize attractively or repulsively coupled scale-free neuronal networks with delays.

\section*{Introduction}
It is well known that synchronization in neuronal
networks is particularly relevant for the efficient processing and
transmission of information (see \textit{e.g.} \cite{up1, up2}).
Experiments have shown that synchronized states can occur in many
special areas of the brain, such as the olfactory system or the
hippocampal region \cite{up3, up4, up5}. By using functional
magnetic resonance imaging (fMRI) to record brain activity from both
speakers and listeners during natural verbal communication, a recent
study has shown that speaker-listener neural coupling underlies
successful communication by means of synchronization \cite{qq1}.
Theoretically, neuronal synchronization on complex networks has been
explored in detail \cite{xn1, xn2, xn3, xn4, WQ, 4, prl}, leading to
several insights that have applicability on real problems in
neuroscience. For example, synchronization of gap-junction-coupled
neurons has been investigated \cite{WQ}, and by means of the phase
resetting curve, phase locking and synchronization in neuronal
networks have been investigated as well \cite{qq2, qq3}. Moreover,
noise-induced and noise-enhanced synchronization have also been
reported in realistic neuronal systems \cite{ni1, ni2}.
Interestingly, it was reported that chemical and electrical synapses
perform complementary roles in the synchronization of interneuronal
networks \cite{CG}. Indeed, synchronization, information
transmission and signal sensitivity on complex networks are
currently hot topics in theoretical neuroscience \cite{r1, r2}, as
evidenced by several recent studies that are devoted to the
explorations of this subject \cite{fh0, fh1, qy1, qy2, fh2, fh3,
fh4, fh5, fh6, fhZ, fhQ, w1}.

Previous research highlighted that information transmission delays
are inherent to the nervous system because of the finite speed at
which action potentials propagate across neuron axons, as well as
due to time lapses occurring by both dendritic and synaptic
processing \cite{or23}. It has been reported, for example, that the
beta frequency is able to synchronize over long conduction delays,
which corresponds to signals traveling a significant distance in the
brain \cite{BD1}. Thus far, it has also been reported that different
time delay lengths can change both qualitative as well as
quantitative properties of the dynamics \cite{AR1}. For example,
delays can introduce or destroy stable oscillations, enhance or
suppress synchronization, as well as generate complex spatiotemporal
patterns on regular neuronal networks. It has also been suggested
that time delays can facilitate neural synchronization and lead to
many interesting and even unexpected phenomena \cite{or24, or25},
including zigzag fronts of excitations, clustering antiphase
synchronization and in-phase synchronization \cite{or26}. Most
recently, the synchronizability threshold for an arbitrary network
incorporating delays and noise has been derived, and additionally,
by means of the scaling theory of the underlying fluctuations, the
absolute limit of synchronization efficiency in a noisy environment
with uniform time delays has been established \cite{prlZ}.

Both phase-attractive (which can be related to excitatory synapses) and phase-repulsive (which can be related to inhibitory synapses) coupling exists in realistic neuronal systems. Hence, it is important to take this explicitly into account in
theoretical studies. Effects of phase-repulsive coupling on
neuronal dynamics have also been investigated in the past
\cite{rep1, rep2, rep3}, where such coupling was considered to be
related to inhibitory synapses. For example, it has been shown that
a pair of excitable FitzHugh-Nagumo neurons can exhibit various
firing patterns including multistability and chaotic firing when
elements interact phase-repulsively \cite{rep1}. Moreover, the
synchronization of nonidentical dynamical units that are coupled
attractively in a small-world network can be improved significantly
by the introduction of just a small fraction of phase-repulsive
couplings \cite{rep3}. Dynamics of propagation in
coupled neuronal networks with excitatory and inhibitory synapses
has been investigated in detail by means of integrate-and-fire neurons
\cite{BD2, BD3}. By analyzing a canard mechanism, it has also been shown that
synaptic coupling can synchronize neurons at low firing
frequencies \cite{BD4}. However, synchronization on scale-free
neuronal networks with phase-repulsive coupling and delay has not
yet been investigated.

Here, we aim to extend the scope of research by
studying the dependence of synchronization transitions on the
information transmission delay over scale-free neuronal networks
with attractive or repulsive coupling, respectively. Since a power
law distribution of the degree of neurons has been found applicable
for the coherence among activated voxels using functional magnetic
resonance imaging \cite{cn2}, and moreover, the robustness against
simulated lesions of anatomic cortical networks was also found to be
most similar to that of a scale-free network \cite{sm2}, our study
addresses a relevant system setup which is still amenable to new
research. We report several non-trivial effects induced by finite
(non-zero) delay lengths, foremost the ability of its fine-tuning
towards highly synchronized fronts of excitations.
We find that the delay-induced synchronization
transitions manifest as well-expressed minima in the
measure for spatiotemporal synchrony. Depending on the type of
coupling, however, these minima appear every integer multiple of the
average oscillation period of bursting oscillations in case of
attractive coupling, or they appear every odd multiple of the
half of the average oscillation period for repulsive coupling.
The results are robust to variations of neuronal
dynamics and system size, and appear to be primarily due to the
emergence of phase locking between the delay and the time scales,
which are inherent to each individual neuron constituting the
scale-free network.

\section*{Results}
\begin{figure}[!ht]
\begin{center}
\includegraphics[width=10cm]{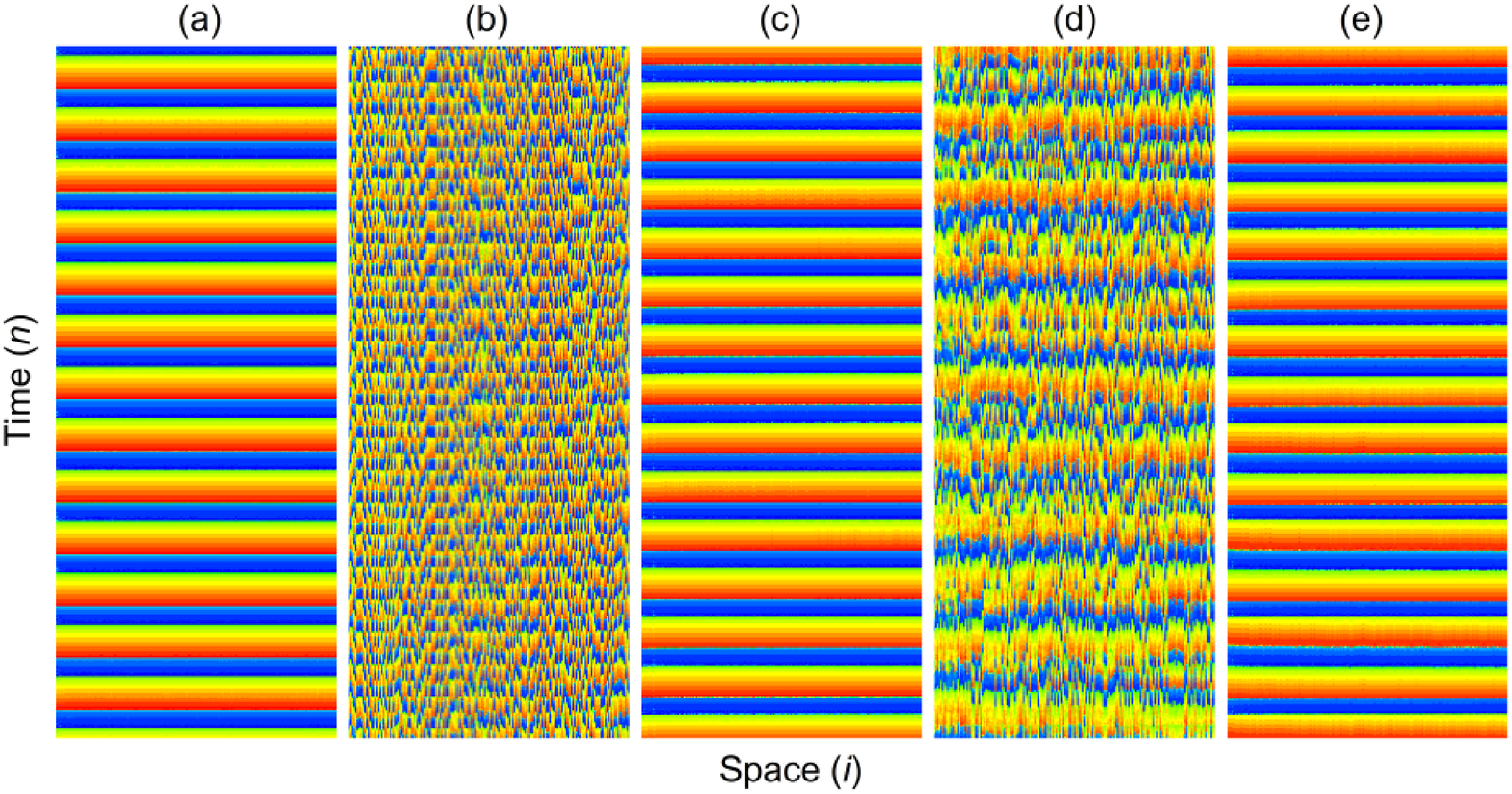}
\end{center}
\caption{\textbf{Characteristic space-time plots of the fast variable $x^{(i)}(n)$ for different information transmission delays $\tau$.} From left to right the delay length is: (a) $0$, (b) $270$, (c) $850$, (d) $1290$ and (e) $1700$. Notice the emergence of complete synchrony in panels (a), (c) and (e). The color coding is linear, red and blue depicting $-2$ and $0.2$ values of $x^{(i)}(n)$, respectively. Other system parameters are: $D=0.01$, $\alpha=2.3$ and $N=200$.}
\label{fig1}
\end{figure}

Firstly, we present in Fig.~\ref{fig1} space-time
plots to have a look at characteristic synchronization transitions
that can be induced by different information transmission delays. To
do so, we employ attractive coupling as a case of example, but note
that qualitatively identical space-time plots can be obtained also
for repulsive coupling. We set $\alpha=2.3$, for which individual
neurons exhibit simple single-burst excitations, as depicted green in Fig.~\ref{fig4}. Results presented in Fig.~\ref{fig1}(a) indicate that the spatiotemporal dynamics is
synchronous if $\tau=0$, which can be attributed to sufficiently
strong attractive coupling. However, if the information transmission
delay is increased to $\tau=270$ the synchrony deteriorates rather
drastically, as can be observed in Fig.~\ref{fig1}(b).
Interestingly, synchronization seems again fully restored at
$\tau=850$, as depicted in Fig.~\ref{fig1}(c), but then again
disappears for $\tau=1290$ and reappears for $\tau=1700$, as shown
in Figs.~\ref{fig1}(d) and (e), respectively. Indeed, we find that
such a succession repeats itself for higher values of $\tau$, from
which we conclude that the information transmission delay can either
promote or impair synchronization of neuronal activity on scale-free
networks. If inspecting the values of $\tau$ warranting near-perfect
synchronization closely, we can observe that they equal roughly
integer multiples of $850$, which hints towards an underlying
mechanism that can explain our observations.

\begin{figure}[!ht]
\begin{center}
\includegraphics[width=11.033cm]{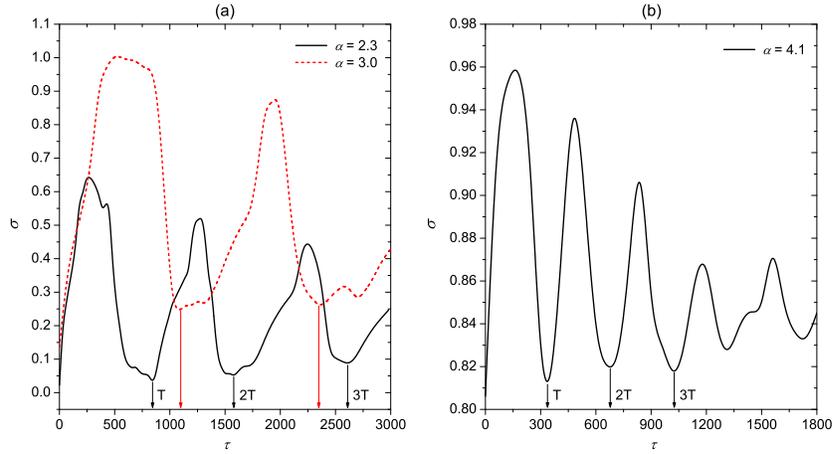}
\end{center}
\caption{\textbf{Quantification of synchronization for attractive
coupling.} Dependence of the synchronization parameter $\sigma$ on
$\tau$ for different values of $\alpha$, as denoted in the
corresponding panels. The undulations of $\sigma$ are clearly
visible and persist irrespective of $\alpha$. While the minima shift
for different $\alpha$, they always occur at integer multiples $T$,
$2T$, $3T$ of the average oscillation period of bursting
oscillations, as denoted by the vertical arrows.}
\label{fig2}
\end{figure}

In order to investigate the impact of different values of $\tau$
quantitatively, and separately for attractive and repulsive
coupling, we calculate the synchronization parameter $\sigma$ as
defined by Eq.~(3). Results presented in Figs.~\ref{fig2}(a) and (b)
were obtained for attractive coupling and three different values of
$\alpha$. It can be observed that certain values of $\tau$
significantly facilitate spatiotemporal synchronization of
excitatory fronts on neuronal scale-free networks. In particular,
the three minima of $\sigma$ appear at $\tau \approx 850=T$, $\tau
\approx 1700=2T$ and $\tau \approx 2550=3T$ if $\alpha=2.3$. For
$\alpha=3.0$, we can observe two minima of $\sigma$ appearing at
$\tau \approx 1200=T$ and $2400=2T$. Furthermore,
several more minima can be observed for $\alpha=4.1$ within the
considered span of information transmission delays, as depicted in
Fig.~\ref{fig2}(b). Again it is clear that they appear at integer
multiples of the first minimum. This confirms the fact that
delay-induced transitions to spatiotemporally synchronized neuronal
activity appear intermittently, at integer multiples of a given
value of $\tau$. On the other hand, values of $\tau$ outside these
regions impair synchronization significantly, as can be inferred
from the rather sharp ascends towards larger values of $\sigma$
beyond the optimal delays.

\begin{figure}
[!ht]
\begin{center}
\includegraphics[width=11.033cm]{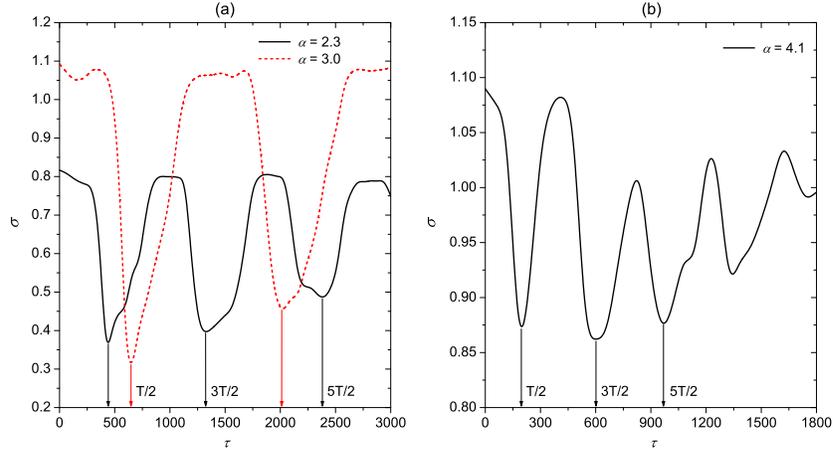}
\end{center}
\caption{\textbf{Quantification of synchronization for repulsive
coupling.} Dependence of the synchronization parameter $\sigma$ on
$\tau$ for different values of $\alpha$, as denoted in the
corresponding panels. The undulations of $\sigma$ are clearly
visible and persist irrespective of $\alpha$. While the minima shift
for different $\alpha$, they always occur at odd integer multiples
$T/2$, $3T/2$, $5T/2$ of the half of the average oscillation period
of bursting oscillations, as denoted by the vertical arrows.}
\label{fig3}
\end{figure}

\begin{figure}[!ht]
\begin{center}
\includegraphics[width=11.033cm]{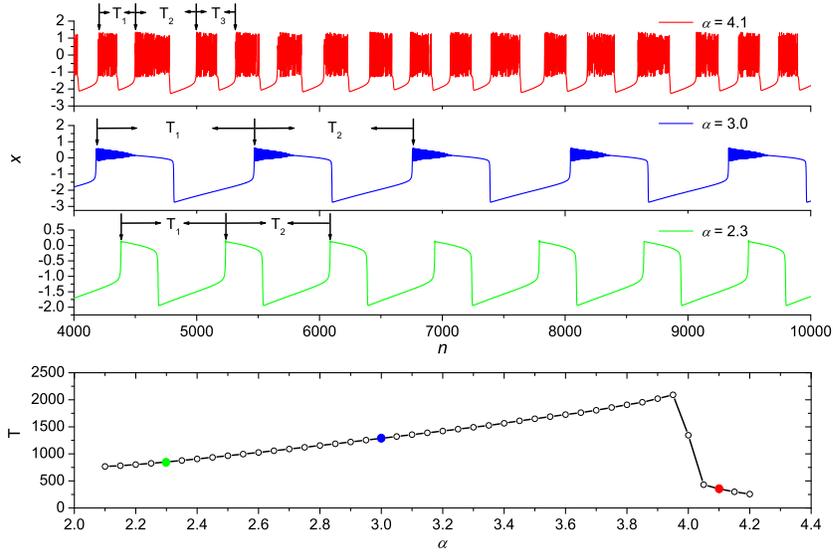}
\end{center}
\caption{\textbf{Time series of the Rulkov map for different values of $\alpha$ and the determination of the average oscillation period of bursting.} \textit{Top three panels:} From top to bottom we have $\alpha=4.1$, $3.0$ and 2.3, respectively. Evidently, the time between consecutive bursts changes significantly, as denoted by $T_{1}$, $T_{2}$ and $T_{3}$, respectively. Simultaneously, different values of $\alpha$ also affect the oscillation period. This gives vital clues as to the location of minima of the synchronization parameter $\sigma$ depicted in Figs.~\ref{fig2} and \ref{fig3}. \textit{Bottom panel:} Average oscillation period of bursting oscillations $T$ in dependence on $\alpha$, determined as the average over $T_{1}, T_{2}, T_{3}, \ldots, T_{L}$. Here $L$ is the total number of periods considered, which was selected large enough to ensure convergence.}
\label{fig4}
\end{figure}

Performing the same analysis for repulsive coupling reveals several
similarities, but also significant differences. Results presented in
Figs.~\ref{fig3}(a) and (b) indeed have a qualitatively identical
outlay with the minima of $\sigma$ appearing intermittently as
$\tau$ increases, yet the precise values warranting optimal neuronal
synchrony are different if compared to the case of attractive
coupling. Specifically, the three minima of $\sigma$ appear at $\tau
\approx 425=T/2$, $\tau \approx 1275=3T/2$ and $\tau \approx
2125=5T/2$ if $\alpha=2.3$, while for $\alpha=3.0$ and $\alpha=4.1$
we can observe similar variations with odd integer
multiples of half of $T$ constituting optimal information
transmission delays where $\sigma$ is minimal. As for attractive
coupling, values of $\tau$ outside these bounds impair
synchronization significantly and fast. Altogether, results
presented in Figs.~\ref{fig2} and \ref{fig3} indicate that simple
scaling laws account for the description of optimal information
transmission delays that warrant near-perfect synchronization of
neuronal activity on scale-free networks. While for attractive
coupling integer multiples of a given constant period are optimal,
for repulsive coupling odd integer multiples of half of the same
period have the best effect. Irrespective of the coupling type,
delays outside the narrow optimal span impair synchronization
significantly.

It is next of interest to explore and determine the mechanisms
behind these observations. We will do this by means of the duration
of bursting periods of individual neurons constituting the
scale-free network. The top three panels of Fig.~\ref{fig4} depict
time courses of the membrane potential $x^{(i)}(n)$ for the values
of $\alpha$ we have used in Figs.~\ref{fig2} and \ref{fig3} above.
It can be observed that, depending on $\alpha$, the duration of
bursts within a given trace may vary (chaotic bursting \cite{rla3}),
but also that the duration of bursts changes due to different
$\alpha$ values. This is highlighted by labels
$T_{1}$, $T_{2}$ and $T_{3}$ (where applicable) in the top three
panels of Fig.~\ref{fig4}. From this it is straightforward to
determine the average oscillation period of bursting $T$ for each
particular value of $\alpha$, simply as the average over a large
enough ensemble $L$ as $T=L^{-1}\sum_{i=1...L}T_{i}$. The bottom
panel of Fig.~\ref{fig4} shows how the average period $T$ varies
with $\alpha$. It can be observed that upon exceeding the Hopf
bifurcation at $\alpha=2.0$ the period $T$ increases fairly
linearly, but then drops rather sharply when $\alpha$ exceeds $4.0$.

Upon connecting the values of $T$ with the optimal information
transmission delays observed in Figs.~\ref{fig2} and \ref{fig3} for
the corresponding values of $\alpha$, we can establish a good
understanding of the mechanism behind the observed synchronization
transitions for attractive as well as for repulsive coupling. In
particular, from results presented in
Fig.~\ref{fig4} it follows that if $\alpha=2.3$ then $T \approx
850$, which is exactly the value of $\tau$ corresponding to the
first minimum of $\sigma$ for attractive coupling. Conversely, one
half and three times one half of $T \approx 850$ correspond to the
first and second minima of $\sigma$ if $\alpha=2.3$ and the coupling
is repulsive. For the other two considered values of $\alpha$,
namely $3.0$ and $4.1$, an identical linkage can be established
easily from the results presented in Figs.~\ref{fig2} and
\ref{fig3}, depending on the type of coupling one is interested in,
and the bottom panel of Fig.~\ref{fig4}. Apparently, the average
period of individual bursts determines the optimal information
transmission delay that warrants the best synchrony, \textit{i.e.}
minimal $\sigma$, of neuronal firings on the scale-free network. We
therefore conclude that for attractive coupling the delay-induced
transitions to spatiotemporal synchronization of neuronal activity
are due to the locking between $\tau$ and the average oscillation
period of individual neurons constituting the scale-free network.
Importantly, because the repulsive coupling can pull adjacent
neurons into antiphase synchronization, the optimal delay warranting
best synchronization is not equal to full integer multiples of $T$.
Thus, it is exactly odd integer multiples of one
half of the average oscillation period of an individual neuron,
where the phase locking between antiphased bursts occurs.

\begin{figure}
[!ht]
\begin{center}
\includegraphics[width=11.033cm]{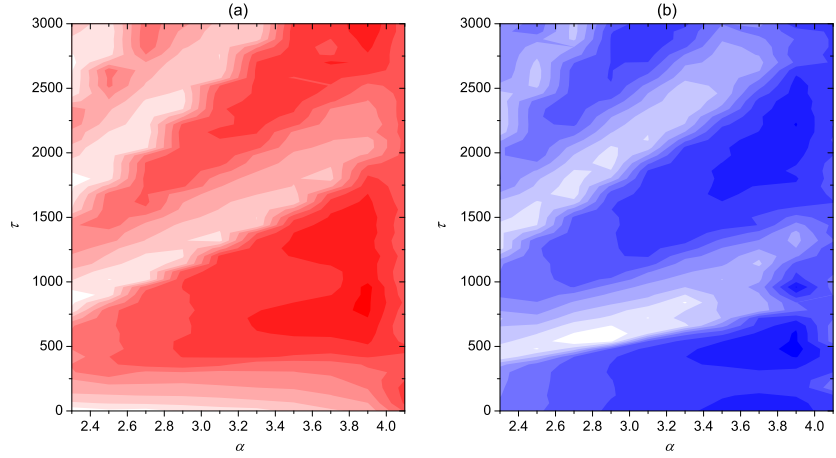}
\end{center}
\caption{\textbf{Two-parameter dependence of synchronization among
neurons.} Contour plots depict the synchronization parameter $\sigma$ in dependence on $\alpha$ and $\tau$ for attractive coupling (panel a) and repulsive coupling
(panel b). Tongues of synchrony (white) emerge due to an intricate
interplay between the inherent dynamics of each neuron constituting
the scale-free network and the locking between the information
transmission delay length and the oscillation period of bursting.}
\label{fig5}
\end{figure}

Merging these observation into an overall insight about
delay-induced synchronization transition on scale-free networks with
attractive and repulsive coupling, we show in
Fig.~\ref{fig5} contour plots of $\sigma$, which depend on the two
main parameters $\alpha$ and $\tau$ for the two types of coupling
separately. The emergence of highly synchronous tongue-like regions
in the two-dimensional parameter plane agrees perfectly with the
reasoning we have outlined above. As the information transmission
delay increases the neuronal activity enters and exits synchronous
regions in an intermittent fashion. Simultaneously, as $\alpha$
increases, the average period of bursting increases nearly linearly
according to the results presented in the bottom panel of
Fig.~\ref{fig4}, thus giving an upward momentum to the white
regions. However, when $\alpha>4.0$ the average oscillation period
drops sharply, which terminates the white ``tongues'' of synchrony
rather abruptly and shifts the optima toward much smaller $\tau$.
Altogether the presented results are in agreement with those
presented in Figs.~\ref{fig2} and \ref{fig3}.

\begin{figure}[!ht]
\begin{center}
\includegraphics[width=11.033cm]{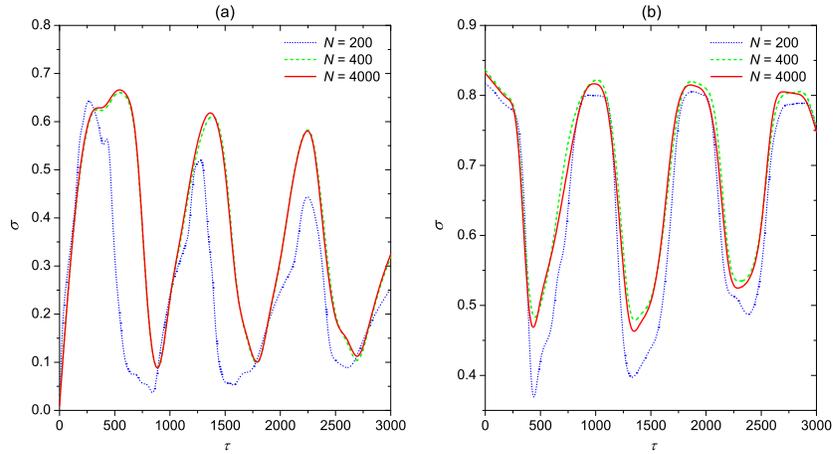}
\end{center}
\caption{\textbf{Dependence of the synchronization parameter
$\sigma$ on $\tau$ for different values of the system size $N$.} (a) Attractive
coupling. (b) Repulsive coupling. Other system parameters
are: $D=0.01$, $\alpha=2.3$. It can be observed that the results vary fairly insignificantly as the system size increases.}
\label{fig6}
\end{figure}

\begin{figure}[!ht]
\begin{center}
\includegraphics[width=11.033cm]{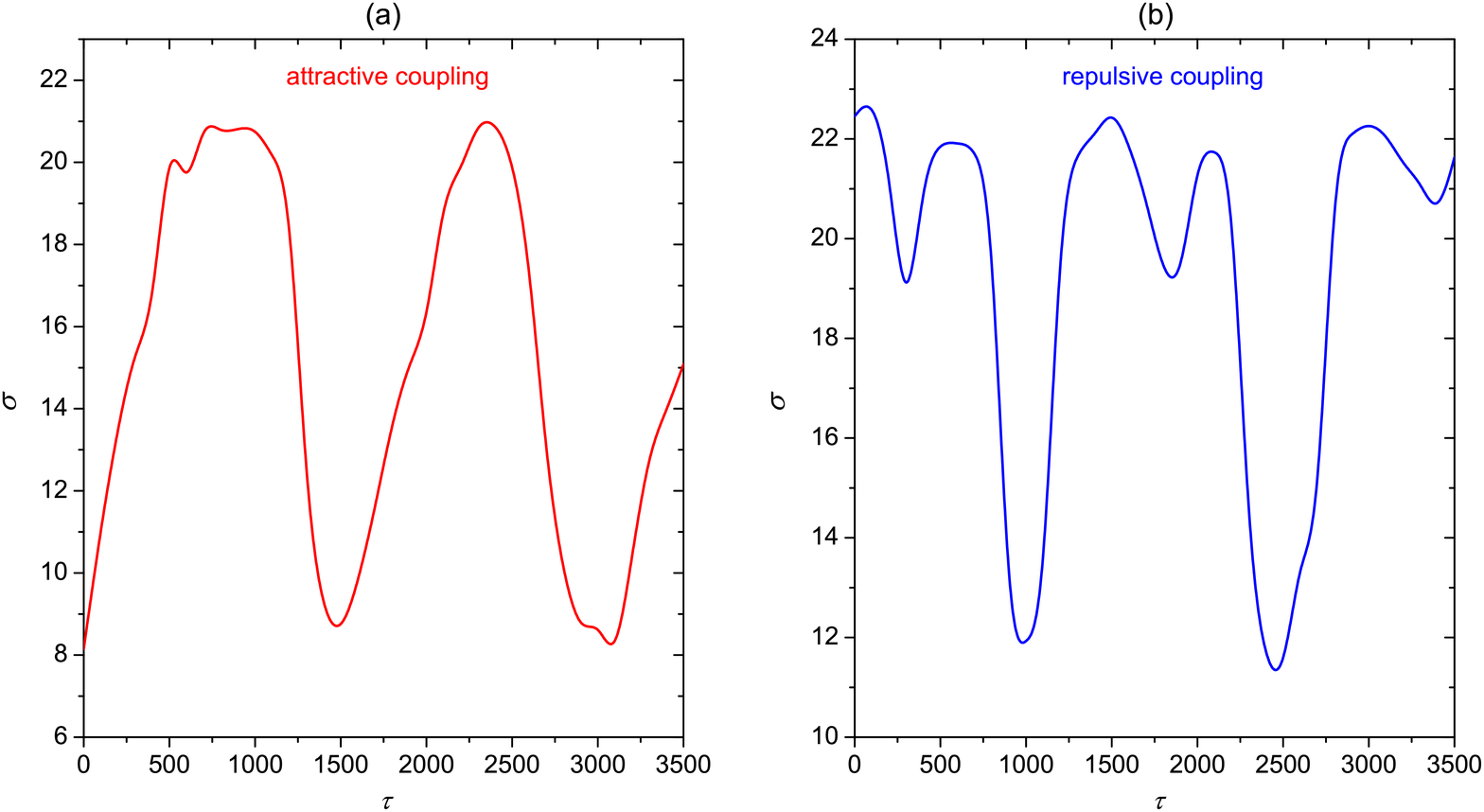}
\end{center}
\caption{\textbf{Dependence of the synchronization parameter $\sigma$ on $\tau$ for type II neuronal dynamics.} (a) Attractive coupling. (b) Repulsive coupling. Other system parameters are: $D=0.02$, $I=10$ and $N=200$. Presented results are qualitatively identical to those obtained with the Rulkov map.}
\label{fig7}
\end{figure}

\begin{figure}[!ht]
\begin{center}
\includegraphics[width=11.033cm]{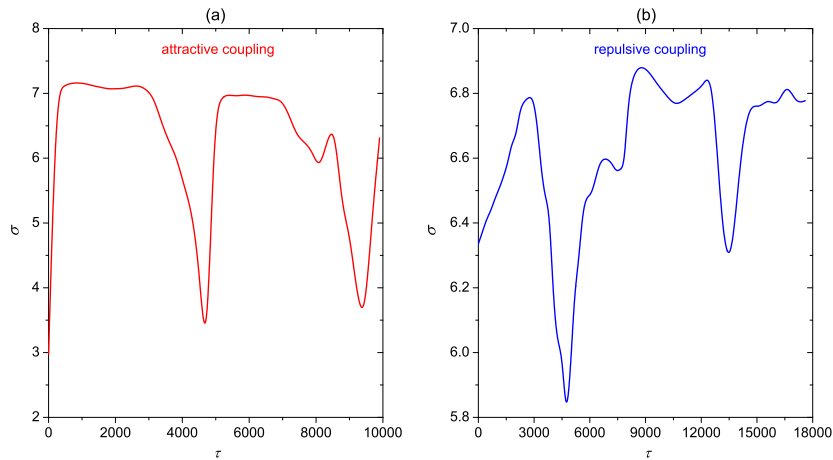}
\end{center}
\caption{\textbf{Dependence of the synchronization parameter $\sigma$ on $\tau$ for type I neuronal dynamics.} (a) Attractive coupling. (b) Repulsive coupling. Other system parameters are: $D=0.01$, $I_{app}=14$ and $N=200$. As in Fig.~\ref{fig7}, the presented results are qualitatively identical to those obtained with the Rulkov map, thus indicating their independence on the particularities of the governing neuronal dynamics.}
\label{fig8}
\end{figure}

In what follows, in order to test the generality of the above results, we investigate the impact of different system sizes $N$ and different models of neuronal dynamics, including those of type I and type II. Firstly, for different system sizes, results depicted in
Figs.~\ref{fig6}(a) and (b) show clearly that the variations of $N$ do not notably influence the outcome of our simulations. In fact, the minima of $\sigma$ remain located at about the same
values of $\tau$ irrespective of $N$. In order to validate our conclusions for different types of neuronal dynamics, we choose the famous Hodgkin-Huxley model (type II) and
the Morris-Lecar model (type I) to describe the dynamics of individual network
nodes (both models are given in the Appendix of the Methods section). Using these two models, we investigate the synchronization transition when the delay
is varied. It is shown in Figs.~\ref{fig7} and \ref{fig8} that irrespectively of the type of the governing neuronal dynamics, intermittent synchronization transitions can still be observed for both the attractive as well as repulsive coupling when the delay is
increased. More importantly, the phase locking between
the delay and the period of oscillators persists in a way that is identical to what we reported above for the Rulkov model. Hence, the obtained results are also
deemed robust against the variations of the neuronal dynamics.

\begin{figure}[!ht]
\begin{center}
\includegraphics[width=11.033cm]{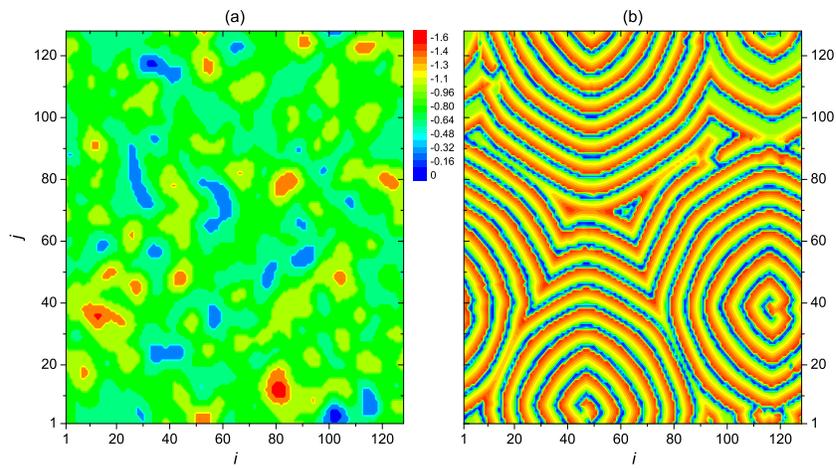}
\end{center}
\caption{\textbf{Delay-induced spatial pattern formation on the square lattice populated by diffusively coupled Rulkov neurons.} Both panels depict values of $x_{i,j}$ on a $128 \times 128$ square lattice at a given (representative) discrete time $n$. The information transmission delay $\tau$ is equal to: (a) $0$, (b) $50$. Coloring in both panels is linear, as depicted by the color strip in the middle, although the scale for the left panel was made much narrower to make the small deviations from the steady state (before it was completely reached) visible. Other system parameters are: $D=0.0025$, $\alpha=1.99$. It can be observed (see panel b) that appropriate information transmission delays evoke ordered excitatory waves with a well-defined spatial frequency.}
\label{fig9}
\end{figure}

Lastly, we construct a square lattice occupying
$128 \times 128$ neurons, whose nodes are modeled by the Rulkov map.
Here we set the parameter $\alpha=1.99$, so that every neuron operates in the excitable regime. Starting with random initial conditions, the results in Fig.~\ref{fig9}(a) evidence that as the delay equals $\tau=0$, there is no pattern formation observable and each neuron approaches its excitable steady state value. On the other hand, however, Fig.~\ref{fig9}(b) features coherent waves of excitation that appear as the delay equals $\tau=50$, which emerge due to the locking between the delay length and the characteristic transient time of the local neuronal dynamics. Hence, it can be concluded that appropriate information transmission delays can also evoke ordered waves of excitation in the spatial domain, thus adding to their importance for the functioning of neuronal tissue.

\section*{Discussion}
We have studied delay-induced synchronization transitions on
attractively and repulsively coupled scale-free neuronal networks
that were locally modeled by the Rulkov map. We have shown that,
irrespective of the type of couplings, information transmission
delays play a pivotal role in ensuring synchronized neuronal
activity. By attractive and repulsive couplings,
the synchronization of bursting oscillations was found undulating
intermittently as the delay was increased. However, while for
attractive coupling the regions of high synchronization appeared
every integer multiple of the average oscillation period, for the
repulsive coupling they appeared every odd multiple of the half of
the average oscillation period. Aiming to explain these observation,
we have argued that by attractive coupling the intermittent outlay
of synchronized regions emerges due to the locking between the delay
length and the average oscillation period of bursting oscillations
of individual neurons constituting the scale-free network.
Conversely, by repulsive coupling the emergence of
antiphase synchronization indicates locking between the delay and
odd multiples of one half of the average oscillation period. Our
results indicate that information transmission delays can either
promote or impair synchrony among neurons and can thus effectively
supplement other mechanisms of synchronization \cite{mh1, mh2} on
scale-free networks, which arguably constitutes an important
ingredient of interneuronal communication. These conclusions seem to
be supported by actual biological data, stating that conduction
velocities along axons connecting neurons vary from 20 to 60 m/s
\cite{TS1}. Real-life transmission delays are thus within the range
of milliseconds, suggesting that substantially lower or higher
values may be preclusive for optimal functioning of neuronal tissue.
Repulsive coupling, as we have considered it in this study, is in
fact an inherent ingredient of several biological systems, in
particular those that contain dynamical units that are in
``competition'' with each other. Known examples are the inhibitory
couplings is present in neuronal circuits associated with a
synchronized behavior in central pattern generators or calcium
oscillations in epileptic human astrocyte cultures \cite{rep3}. We
hope that these results will foster our understanding of the
observed neuronal activity.

\section*{Methods}
The map proposed by Rulkov \cite{r20, rla4} determines the dynamics
of individual nodes forming the scale-free network. It captures
succinctly the main dynamical features of the more complex
time-continuous neuronal models, but simultaneously allows an
efficient numerical treatment of large systems \cite{rla1}.
Accordingly, the spatiotemporal evolution of the studied network
with information transmission delay is governed by the following
iteration equations
\begin{eqnarray}
x^{(i)}(n+1)&=& \alpha f[x^{(i)}(n)]+y^{(i)}(n)+ D\sum_{j}\varepsilon^{{i,j}}\left[x^{j}(n-\tau)-x^{i}(n)\right],\\
y^{(i)}(n+1)&=& y^{(i)}(n)-\beta x^{(i)}(n)-\gamma, \ i=1,\ldots,N,
\end{eqnarray}
where $f(x)=\frac{1}{1+x^2}$ is a nonlinear function warranting the essential ingredients of neuronal dynamics, $x^{(i)}(n)$ is the membrane potential of the $i$-th neuron and $y^{(i)}(n)$ is the variation of the ion concentration, the two representing the fast and the slow variable of the map, respectively. The slow temporal evolution of $y^{(i)}(n)$ is due to the small values of the two parameters $\beta$ and $\gamma$ that are here both set equal to $0.001$. Moreover, $n$ is the discrete time index, while $\alpha$ is the main bifurcation parameter determining the dynamics of individual neurons constituting the scale-free network. In \cite{r20} it was shown that for $\alpha < 2.0$ all neurons are situated in excitable steady states $[x^* = -1,y^* = -1-(\alpha/2)]$, whereas if $\alpha > 2.0$ complex oscillatory and bursting patterns can emerge via a Hopf bifurcation. Importantly, we set the coupling strength equal to either $D=0.01$, corresponding to attractive coupling, or $D=-0.01$, corresponding to repulsive coupling. Parameter $\tau$ is the information transmission delay that together with $\alpha$ represents the two crucial parameters that are varied in the realm of this study.

As the interaction network between neurons we use the scale-free
network generated via growth and preferential attachment as proposed
by Barab\'{a}si and Albert \cite{r21}, typically consisting of
$N=200$ nodes or more. Each node corresponds to one neuron, whose dynamics
is governed by the Rulkov map, as described above. In Eq.~(1)
$\varepsilon^{i,j} = 1$ if neuron $i$ is coupled to neuron $j$ and
$\varepsilon^{i,j} = 0$ otherwise. Following \cite{r21}, the
preferential attachment is introduced via the probability $\Pi$,
which states that a new node will be connected to node $i$ depending
on its connectivity $k_i$ according to $\Pi(k_i )=k_i /\sum_jk_j$.
Here, $k_i$ is the degree of node $i$ (the degree
of a node is the number of links adjacent to it). This growth and
preferential attachment scheme yields a network with an average
degree $k_{av}=\frac{\sum_i k_i}{N}$, and a power-law degree
distribution with the slope of the line equaling $\approx -3$ on a
double-logarithmic graph. We will use Barab\'{a}si-Albert scale-free
networks having $k_{av}=4$ throughout this work.

In order to study synchronization transitions quantitatively, we introduce, by means of the standard deviation, a synchronization parameter $\sigma$ (see \textit{e.g.} \cite{r22}), which can be calculated effectively according to:
\begin{equation}
\sigma=\sqrt{\frac{1}{T}\sum\limits_{n=1}^{T}\sigma(n)}, \ \
\sigma(n)=\frac{1}{N}\sum\limits_{i=1}^{N}[x^{i}(n)]^2-\left[\frac{1}{N}\sum\limits_{i=1}^{N}x^{i}(n)\right]^2.
\end{equation}
In particular, $\sigma$ is an excellent quantity for numerically effectively measuring the spatiotemporal synchronization of excitations, hence revealing different synchronization levels and with it related transitions. From Eq.~(3) it is evident that the more synchronous the neuronal network the smaller the synchronization parameter $\sigma$. Accordingly, in the event of complete synchrony we have $\sigma=0$. Presented results were averaged over $20$ independent runs for each set of parameter values to warrant appropriate statistical accuracy with respect to the scale-free network generation and numerical simulations.

\subsection*{Appendix: Alternative models of neuronal dynamics}
The full Hodgkin-Huxley model is given by the following equations \cite{HH52}:
\begin{eqnarray}
\label{eq:eq1} C \frac{{\rm d}V}{{\rm d}t}&=&-g_{Na}m^3h(V-V_{Na})-
g_{L}(V-V_{L})-g_{K}X_{K}n^4(V-V_{K})+I, \nonumber \\
\frac{{\rm d}m}{{\rm d}t}&=&\alpha_{m}(1-m)-\beta_{m}{m},\nonumber \\
\frac{{\rm d}h}{{\rm d}t}&=&\alpha_{h}(1-h)-\beta_{h}{h},\nonumber \\
\frac{{\rm d}n}{{\rm d}t}&=&\alpha_{n}(1-n)-\beta_{n}{n},\nonumber
\end{eqnarray}
where $V$ is the transmembrane potential of the neuron, and $m$,
$h$ and $n$ are the corresponding gating variables (probabilities)
characterized by a two-state, opening or closing dynamics. The
voltage-dependent opening and closing rates are given explicitly by
the following expressions:
\begin{eqnarray}
\alpha_{m}&=&\frac{0.1(V+10)}{1-{\rm exp}[-\frac{(V+40)}{10}]},\nonumber \\
\beta_{m}&=&4 {\rm exp}\left[-\frac{(V+65)}{18} \right], \nonumber \\
\alpha_{h}&=&0.07 {\rm exp}\left[-\frac{(V+65)}{20}\right], \nonumber \\
\beta_{h}&=&\left\{1+{\rm exp}\left[-\frac{(V+35)}{10}\right]\right\}^{-1},\nonumber \\
\alpha_{n}&=&\frac{0.01(V+55)}{1-{\rm exp}[-\frac{(V+55)}{10}]}, \nonumber \\
\beta_{n}&=&0.125 {\rm exp}\left[-\frac{(V+65)}{80}\right].\nonumber
\end{eqnarray}
The membrane capacity $C$=1 ($\mu$F/cm$^{2}$), parameters $g_{Na}$,
$g_{K}$ and $g_{L}$ are maximal sodium, potassium and leakage
conductances, $g_{Na}=120$ $\mu$F/cm$^{2}$, $g_{K}=36$
$\mu$F/cm$^{2}$ and $g_{L}=0.3$ $\mu$F/cm$^{2}$, respectively,
$V_{Na}$, $V_{K}$, and $V_{L}$ are the reversal potentials,
$V_{Na}=50$ mV, $V_{K}=-77$ mV, $V_{L}=-54.4$ mV and $I=10$ $\mu$.

The dynamics of the type I Morris-Lecar neuron is described by the
following equations \cite{ML}:
\begin{eqnarray}
C\frac{dV}{dt}&=-&I_{Ca}-I_{K}-I_{L}+I_{app}, \nonumber \\
\frac{d\omega}{dt}&=&\frac{(\omega_{\infty}-\omega)}{\tau_{\infty}(V)},\nonumber\\
I_{Ca}&=& g_{Ca}m_\infty(V)(V-V_{Ca}),\nonumber \\
I_{K}&=& g_{K}\omega(V)(V-V_{K}),\nonumber\\
I_{L}&=& g_{L}(V-V_{L}),\nonumber
\end{eqnarray}
where $V$ is the cell membrane potential in mV, $I_{Ca}$ is the
depolarizing calcium current, $I_L$ is the passive leak current,
respectively, $\omega$ is the activation of the repolarizing
potassium current $I_{K}$, $t$ is time in ms, and $Iapp =14$ $\mu$
A/cm$^{2}$ is the applied current. The remaining parameters are
$V_{Ca}$ =120 mV, $V_K =-84$ mV, $V_L =-60$ mV, $g_{Ca}$
=4mS/cm$^{2}$, $g_K$ = 8 mS/cm$^{2}$, $g_L$ = 2mS/cm$^{2}$. The steady state activation of the calcium current is:
\begin{eqnarray}
m_\infty(V)=\frac{1}{2}\left[1+\tanh\left(\frac{V+12}{18}\right)\right].
\nonumber
\end{eqnarray}
The potassium current activation amplitude and activation rate are:
\begin{eqnarray}
\omega_\infty(V)&=&\frac{1}{2}\left[1+\tanh\left(\frac{V+8}{6}\right)\right], \nonumber \\
\frac{1}{\tau_\infty(V)}&=&\frac{2}{3}\cosh\left(\frac{V+12}{18}\right).
\nonumber
\end{eqnarray}

\clearpage

\end{document}